# On Enforcing Existence and Non-Existence Constraints in *MatBase*

**Christian Mancas***

*Department of Mathematics and Computer Science, Ovidius University, Constanta, Romania*

**\*Corresponding Author:** Christian Mancas, Department of Mathematics and Computer Science, Ovidius University, Constanta, Romania.

**Abstract**

Existence constraints were defined in the Relational Data Model, but, unfortunately, are not provided by any Relational Database Management System, except for their NOT NULL particular case. Our (Elementary) Mathematical Data Model extended them to function products and introduced their dual non-existence constraints. *MatBase*, an intelligent data and knowledge base management system prototype based on both these data models, not only provides existence and non-existence constraints, but also automatically generates code for their enforcement. This paper presents and discusses the algorithms used by *MatBase* to enforce these types of constraints.

## Introduction

In the Relational Data Model (RDM) [1-3], an *existence constraint* is denoted f |— g, where *f* and *g* are distinct no not-null columns of a same table *R*, asks for *g* to have a not-null value whenever *f* has a not-null value in the corresponding database (db) instance. For example, in a *PERSONS* table, *SSN* |— *BirthDate* asks that *BirthDate* be not-null whenever *SSN* is not-null (i.e., for any person, whenever his/her social security number is stored in the db, his/her birth date must also be stored).

Unfortunately, no commercial Relational Database Management System (RDBMS) is providing this constraint type, but only its particularization $\varnothing$ |— *g*, which is called *NOT NULL* (or *required values*) *constraint*.

You cannot use the *tuple constraint type* (i.e., constraints of the type $(\forall x \in D)(P(x))$, where *P* is a first order logic predicate involving only functions defined on *D*) either for enforcing existence constraints in commercial RDBMSes. For example, MS SQL Server syntactically accepts a check constraint reading "not *f* is null and not *g* is null" for a table containing columns *f* and *g* accepting nulls (and, e.g., both of type int), but you cannot save it in the schema of that table (e.g., under the name "CK_test"), as it is semantically rejected with the following error message: "Unable to add constraint 'CK_test'. The ALTER TABLE statement conflicted with the CHECK constraint 'CK_test'".

In our (Elementary) Mathematical Data Model ((E)MDM) [4, 5], tables are considered (object) sets and their columns are functions defined on the corresponding sets and taking values in either value (i.e., data type) or object sets. Hence, the NOT NULL constraints are called *total* ones (i.e., corresponding functions are totally defined). Besides $\emptyset$ (i.e., the empty set), (E)MDM also considers the distinguished countable set of *null values* (be them temporarily unknown or inapplicable), denoted *NULLS*. For example, *SSN* • *BirthDate* : *PERSONS* → (NAT(9) ∪ NULLS) x ([1/1/1900, *Today*()] ∪ NULLS) may be simply declared as *SSN* • *BirthDate* : *PERSONS* → NAT(9) x [1/1/1900, *Today*()], as none of these two functions is totally defined (because not everybody is entitled to an SSN).

(E)MDM also extends the existence constraints, by allowing on both sides of the implication sign, computed functions, not only atomic ones. For example, *SSN* |— *BirthDate* • *Sex* asks that both *BirthDate* and *Sex* be not-null whenever *SSN* is not-null, while *EYear* |— *FName* ° *FirstAuthor* ° *FirstBook*, forces users to also specify the first name for the first author of any book for which there is at least an edition having that book as its first one and for which the edition year is known (see the *Public Library* db from [3]). Generally, for any $f \bullet g : D \to C_f \times C_g$, $f$ |— $g$, means (($\forall x \in D$) ($f(x) \notin$ NULLS $\Rightarrow g(x) \notin$ NULLS).

Moreover, (E)MDM also considers a dual of existence constraints: a *non-existence constraint* is denoted $f \neg$ |— $f_1 \bullet \ldots \bullet f_n$ and stands for ($\forall x \in D$)($f(x) \notin$ NULLS $\Rightarrow f_i(x) \in$ NULLS, $\forall i, 1 \leq i \leq n$), where $n > 0$ is a natural. For example, *TributaryTo* $\neg$ |— *Lake* • *Sea* • *Ocean* • *LostInto* asks that, whenever a river is tributary to another one, its values for *Lake*, *Sea*, *Ocean*, and *LostInto* be nulls.

For cases in which at most one function from a product might take a not null value, (E)MDM provides the abbreviation $\neg$ |— $f_1 \bullet \ldots \bullet f_n$, which stands for ($\forall x \in D$)($f_j(x) \notin$ NULLS $\Rightarrow f_i(x) \in$ NULLS, $\forall i, 1 \leq i \neq j \leq n$), where $n > 1$ is a natural. For example, $\neg$ |— *TributaryTo* • *Lake* • *Sea* • *Ocean* • *LostInto* formalizes the constraint that a river may be either tributary to another one or end in a lake, or sea, or ocean, or be lost (e.g., in a desert, cave, etc.), which consolidates five atomic non-existence constraints.

Currently, such constraints must be manually enforced by software developers of db software applications. Our *MatBase* [4-6], an intelligent data and knowledge base management system prototype based on both (E)MDM, RDM, the Entity-Relationship Data Model [3, 7, 8], and Datalog¬ [2, 5], provides in its (E)MDM interface a Graphic User Interface (GUI) MS Windows form for declaring both existence and non-existence constraints; if they are accepted, *MatBase* is automatically generating code for their enforcement. This paper presents and discusses the corresponding algorithms, which, of course, may also be used by any developer not having access to a copy of *MatBase*.

*Literature Review*

Existence constraints were defined in the framework of RDM [3, 4, 5, 9]. Non-existence constraints were introduced by us [4, 5]. Generally, existence constraints were very rarely studied.

In the ontology realm, RDM existence constraints are semantically divided into three subtypes (namely, *exclusive*, for not-applicable nulls, *mutual*, when $f$ and $g$ have same applicability domains, and *conditioned*, for the original RDM concept) [10].

As frequently happens, other concepts are also known in the literature under the "existence constraint" name. For example, [11] uses it in the context of dynamic (transactional) integrity constraints; in the NoSQL Memgraph and Neo4j DBMSes, the NOT NULL constraints are called existence constraints [12, 13]; existence constraints are also used with a completely different meaning (i.e., they "assert that for all (medial-level) variable values from a set of infinite cardinality, there must exist (lower-level) variable values from a second set that satisfy an inequality") in the realm of semi-infinite programs [14].

## Methodology

*Well-formed existence and non-existence constraints*

First, please note that only functions which are not totally defined may be involved in existence and non-existence constraints. Indeed, if, for example, $f$ is totally defined, then $f$ |— $g$ implies that $g$ must also be totally defined (as $f$ does not ever take null values); dually, if $g$ is totally defined, then $f$ |— $g$ is superfluous (as g does not ever take null values). For the dually non-existence constraints

things are similar. Indeed, if, for example, *f* is totally defined, then $f \neg |\!\!-\!\!- g$ implies that g must always be null, which is absurd: why would you add to a table a column that is allowed to store nothing? Conversely, if *g* is totally defined, then $f \neg |\!\!-\!\!- g$ is unenforceable, except when *f* is always taking null values.

In terms of incoherence (i.e., no function image should always be the empty set or a subset of NULLS) and redundancy (i.e., no constraint should be implied by others) of constraint sets, this may be summarized in the following proposition:

Proposition 1.

(i) $(f |\!\!-\!\!- g \wedge f$ total $\Rightarrow g$ total$) \Rightarrow f |\!\!-\!\!- g$ redundant.
(ii) $f |\!\!-\!\!- g \wedge g$ total $\Rightarrow f |\!\!-\!\!- g$ redundant.
(iii) $(f \neg |\!\!-\!\!- g \wedge f$ total $\Rightarrow \forall x, g(x) \in$ NULLS$) \Rightarrow \{f \neg |\!\!-\!\!- g, f$ total$\}$ incoherent.
(iv) $(f \neg |\!\!-\!\!- g \wedge g$ total $\Rightarrow \forall x, f(x) \in$ NULLS$) \Rightarrow \{f \neg |\!\!-\!\!- g, g$ total$\}$ incoherent.

Consequently, any existence or non-existence constraint must not be accepted if at least one of the functions implied is totally defined.

Secondly, which is obvious according to their definitions, both existence and non-existence constraints are subtypes of the tuple constraints.

Consequently, any existence or non-existence constraint must not be accepted if at least one of the functions implied is not defined on a subset of a set that includes the domains of the rest of the implied functions. This restriction includes the existence and non-existence constraints in the subcategory of constraints on function products.

For example, $\neg |\!\!-\!\!-$ *SSN • ITIN* (where *ITIN* stands for the Individual Taxpayer Identification Number) is well-formed as both are defined on subsets of the *USResidences* set, a subset of *PERSONS*. On the contrary, *BirthPlace* $|\!\!-\!\!-$ *Country* is not a valid existence constraint, as *BirthPlace* : *PERSONS* → *CITIES*, while *Country* : *CITIES* → *COUNTRIES*.

To conclude with, a *well-formed existence or non-existence constraint* must be defined over a function product whose components are not totally defined functions.

Figure 1 shows *MatBase's* pseudocode algorithm for accepting, rejecting, and deleting existence and non-existence constraints.

*Enforcing existence and non-existence constraints*

First, please note that, obviously, deleting an element from the underlying set of an existence or non-existence constraint, i.e., a line from the corresponding db table, cannot violate any such constraint.

On the contrary, both adding a new element (line) and modifying an existent one might violate such constraints. Let us consider $f = f_1 \bullet \ldots \bullet f_n : D_f \to C_f$, $g = g_1 \bullet \ldots \bullet g_m : D_g \to C_g$, $n, m$ naturals, $D_f \subseteq D$, and $D_g \subseteq D$.

According to its definition, the existence constraint $f |\!\!-\!\!- g$ is violated for a $x \in D$ whenever $f_i(x) \notin$ NULLS, $1 \leq i \leq n$, and there is at least a *j* in $1 \leq j \leq m$ such that $g_j(x) \in$ NULLS.

According to its definition, the non-existence constraint $f \neg |\!\!-\!\!- g$ is violated for a $x \in D$ whenever $f_i(x) \notin$ NULLS, $1 \leq i \leq n$, and there is at least a *j* in $1 \leq j \leq m$ such that $g_j(x) \notin$ NULLS, whereas a consolidated set of non-existence constraints of the type $\neg |\!\!-\!\!- g$ is violated for a $x \in D$ whenever there are at least two values $i, j$ in $1 \leq i, j \leq m$, $i \neq j$, such that $g_i(x) \notin$ NULLS and $g_j(x) \notin$ NULLS.

```
Algorithm A1: adding or deleting existence and non-existence constraints
Input: the current constraint set C, the required operation type ot (true for add or false for delete),
       constraint name cn, constraint type ct (true for existence or false for non-existence), left-side
       function f = f₁ • … • fₙ : D_f → C_f, and right-side function g = g₁ • … • g_m : D_g → C_g, n, m
       naturals;
Output: C — {cn}, if ¬ct and cn's deletion is confirmed or C ∪ {cn}, if cn is admissible;
Strategy:
if ¬ct then if cn ∈ C then if user confirms his/her request to delete cn then C = C — {cn};
            else display "Request rejected: " & cn & " is not a known constraint name!";
else if there is a constraint in C named cn then display "Request rejected: " & cn & " is the name " &
                    "of another constraint! Please choose a unique constraint name instead!";
     else if ct and n == 0 then display "Request rejected: please add to C the constraint " & g &
                    " total instead!";
          else if there is no set D such that D_f ⊆ D and D_g ⊆ D then display "Request rejected :" & f
                    & " and " & g & " do not have compatible domains!";
               else Boolean ok = true; int i = 1;
                    while ok and i ≤ n
                       if f_i total then ok = false; display "Request rejected: " & f_i & " is totally defined!";
                       else i = i + 1;
                    end while;
                    if ok then i = 1;
                       while ok and i ≤ m
                          if g_i total then ok = false; display "Request rejected: " & g_i & " is totally defined!";
                          else i = i + 1;
                       end while;
                       if ok then if there is x ∈ D such that cn is violated for it then
                                display "Request rejected: " & cn & " is violated for " & x & "!";
                             else C = C ∪ {cn};
                    end if;
               end if;
          end if;
     end if;
end if;
End Algorithm A1;
```

***Figure 1:*** *MatBase* algorithm for adding and deleting existence and non-existence constraints.

Consequently, to enforce such constraints the class *D* of the software application managing the db to which *D* belongs must contain an event-driven method *BeforeUpdate* like the one shown in Figure 2 that must be automatically launched whenever the current line from table *D* (uniquely identified by the value of *D*'s surrogate primary key *x*) has been modified and the user asks for saving the modifications into the corresponding db. If such a method does not exist, then *MatBase* automatically generates it the first time that a constraint involving *D* is added to the db scheme.

```
method BeforeUpdate()
Boolean Cancel = false;
if not Cancel then Cancel = enforce_non-existence_cnstr(nec, f(x), g(x));
if not Cancel then Cancel = enforce_existence_cnstr(ec, f(x), g(x));
if Cancel then reject saving the current line into the db;
end method BeforeUpdate;
```

***Figure 2:*** The event-driven method *BeforeUpdate* of a class *D* associated to a table *D*.

Whenever *MatBase* accepts an existence constraint *ec*: *f* |—*g* over *D*, it adds to this method a line that reads "*if not Cancel then Cancel = enforce_existence_cnstr*(*ec*, *f*(*x*), *g*(*x*));*"; whenever *MatBase* accepts a non-existence constraint *nec*: *f* ¬|—*g* over *D*, it adds a line that reads "*if not Cancel then Cancel = enforce_non-existence_cnstr*(*nec*, *f*(*x*), *g*(*x*));*". All lines are added to this method immediately after the line that reads "Boolean *Cancel* = *false*;". This is how *MatBase* implements the *C* = *C* ∪ {*cn*} statement of the Algorithm *A*1 from Figure 1.

Dually, whenever *MatBase* deletes an existence constraint named *ec* over *D*, it deletes from this method the line that starts with "*if not Cancel then Cancel = enforce_existence_cnstr*(*ec*,"; whenever *MatBase* deletes a non-existence constraint named *nec* over *D*, it deletes the line that starts with "*if not Cancel then Cancel = enforce_non-existence_cnstr*(*nec*,". This is how *MatBase* implements the *C* = *C* — {*cn*} statement of the Algorithm *A*1 from Figure 1.

Figures 3 and 4 present the pseudocode of methods *enforce_existence_cnstr* and *enforce_non-existence_cnstr*, respectively, that *MatBase* is storing in one of its libraries.

```
Boolean method enforce_existence_cnstr(ec, f(x), g(x));
// returns true if the values of f(x) and g(x) violate ec and false otherwise
enforce_existence_cnstr = false;
if x is a new line or f(x) has been modified or g(x) has been modified then
    Boolean null = true; int i = 1;
    while null and i ≤ n
        if f_i(x) ∉ NULLS then null = false; else i = i + 1;
    end while;
    if not null then i = 1;
        while not null and i ≤ m
            if g_i(x) ∈ NULLS then null = true; else i = i + 1;
        end while;
        if null then enforce_existence_cnstr = true; display "Saving these values is rejected: according"
                & " to existence constraint " & ec & ", column " & g_i & " must have a not null value!"
    end if;
end if;
end method enforce_existence_cnstr;
```

*Figure 3:* *MatBase* algorithm for enforcing existence constraints.

## Results and Discussions

Analyzing Algorithm *A*1, it is easy to prove the following proposition:

*Proposition* 2: Algorithm *A*1 from Figure 1 has the following properties:

(i) It never loops forever.
(ii) Its complexity is $O(n + m)$.
(iii) It deletes existence and non-existence constraints, if the corresponding requests are confirmed, without any check, as such a deletion cannot affect either the coherence or the minimality of the remaining constraint set (recall that a set of constraints is *minimal* if and only if it does not include any redundant constraint).
(iv) It enforces the unicity of the constraint names.
(v) It rejects all totality constraints, i.e., particular cases of existence constraints of type ∅ |—*f*.
(vi) It only adds well-formed existence or non-existence constraints.
(vii) It adds existence and non-existence constraints only if the current db instance does not violate them.
(viii) It is optimal, i.e., it performs its tasks with the minimum possible number of statements.

Analyzing method *BeforeUpdate*, it is easy to prove the following proposition:

*Proposition* 3: Method *BeforeUpdate* from Figure 2 has the following properties:

   (i)     It never loops forever.
   (ii)    Its complexity is $k = 4$.
   (iii)   It enforces any valid existence and non-existence constraints.
   (iv)    It ends as soon as it discovers that a constraint is violated.
   (v)     It is optimal, i.e., it performs its tasks with the minimum possible number of statements.

Analyzing method *enforce_existence_cnstr*, it is easy to prove the following proposition:

Proposition 4: Method *enforce_existence_cnstr* from Figure 3 has the following properties:

   (i)     It never loops forever.
   (ii)    Its complexity is $O(n + m)$.
   (iii)   It returns *true* if the values of $f(x)$ and $g(x)$ violate *ec* and *false* otherwise.
   (iv)    It does not check *ec* if neither $f(x)$ nor $g(x)$ were updated.
   (v)     It does not even check $g(x)$ if $f(x)$ only has null values; otherwise, it ends as soon as it discovers a null value in $g(x)$.
   (vi)    Whenever *ec* is violated, it displays a corresponding error message, also indicating the value that violates it.
   (vii)   It is optimal, i.e., it performs its tasks with the minimum possible number of statements.

```
Boolean method enforce_non-existence_cnstr(nec, f(x), g(x));
// returns true if the values of f(x) and g(x) violate nec and false otherwise
enforce_non-existence_cnstr = false;
if x is a new line or f(x) has been modified or g(x) has been modified then
   int i = 1;
   if f(x) == ∅ then        // consolidated set of non-existent constraints
      int j = 0;            // position of the first not null value in g(x)
      while not enforce_non-existence_cnstr and i ≤ m
         if g_i(x) ∉ NULLS then
            if j == 0 then j = i;
            else enforce_non-existence_cnstr = true;
               display "Saving these values is rejected: according to non-existence constraint " &
                       nec & ", only one of the columns " & g_i & " and " & g_j & " may have a " &
                       "not null value!";
            end if;
         else i = i + 1;
      end while;
   else                     // single non-existent constraint
      Boolean null = true;
      while null and i ≤ n
         if f_i(x) ∉ NULLS then null = false; else i = i + 1;
      end while;
      if not null then i = 1;
         while not null and i ≤ m
            if g_i(x) ∉ NULLS then null = true; else i = i + 1;
         end while;
         if null then enforce_non-existence_cnstr = true;
            display "Saving these values is rejected: according to non-existence constraint " &
                    nec & ", column " & g_i & " must have a null value!"
      end if;
   end if;
end if;
end method enforce_non-existence_cnstr;
```

**Figure 4:** *MatBase* algorithm for enforcing non-existence constraints.

Analyzing method *enforce_non-existence_cnstr*, it is easy to prove the following proposition:

*Proposition* 5: Method *enforce_non-existence_cnstr* from Figure 4 has the following properties:

(i) It never loops forever.
(ii) Its complexity is $O(n + m)$.
(iii) It returns *true* if the values of $f(x)$ and $g(x)$ violate *nec* and *false* otherwise.
(iv) It does not check *nec* if neither $f(x)$ nor $g(x)$ were updated.
(v) It correctly distinguishes between single non-existent constraints and consolidated set of such constraints.
(vi) For any consolidated set of non-existent constraints, it ends as soon as it discovers a second not null value in $g(x)$.
(vii) For single non-existent constraints, it does not even check $g(x)$ if $f(x)$ only has null values; otherwise, it ends as soon as it discovers a not null value in $g(x)$.
(viii) Whenever *nec* is violated, it displays a corresponding error message, also indicating the value that violates it.
(ix) It is optimal, i.e., it performs its tasks with the minimum possible number of statements.

For example, let us consider *D* = PERSONS, *ec*: SSN • ITIN |— BirthDate • Sex, and *nec*: ¬|—SSN • ITIN;

1. Obviously, both *ec* and *nec* are accepted by Algorithm *A*1, so that the *BeforeUpdate* method of class *PERSONS* would look exactly as in Figure 5.
2. Any attempt to save to the db a line {(SSN • ITIN)(x) = <123456789,>, (BirthDate • Sex)(x) = < , "F">} is rejected with the error message "Saving these values is rejected: according to existence constraint *ec*, column *BirthDate* must have a not null value!".
3. Any attempt to save to the db a line {(SSN • ITIN)(x) = <123456789,>, (BirthDate • Sex)(x) = < "1/1/1990", >} is rejected with the error message "Saving these values is rejected: according to existence constraint *ec*, column *Sex* must have a not null value!".
4. Any attempt to save to the db a line { , (SSN • ITIN)(x) = <123456789, 987654321>} is rejected with the error message "Saving these values is rejected: according to non-existence constraint *nec*, column *ITIN* must have a null value!".

```
method BeforeUpdate()
Boolean Cancel = false;
if not Cancel then Cancel = enforce_non-existence_cnstr(nec, , (SSN • ITIN)(x));
if not Cancel then Cancel = enforce_existence_cnstr(ec, (SSN • ITIN)(x), (BirthDate • Sex)(x));
if Cancel then reject saving the current line into the db;
end method BeforeUpdate;
```

*Figure 5:* The event-driven method *BeforeUpdate* of class *PERSONS* associated to table *PERSONS*.

## Conclusion

To guarantee data plausibility, the highest possible quality standard in db software applications, all business rules governing corresponding subuniverses of interest must be formalized as constraints and enforced either by the underlying DBMSes or/and by the software applications. Existence constraints abound in any such subuniverse and were formalized in the RDM framework. However, commercial RDBMSes only provide their NOT NULL particular case, leaving the enforcement of the general case to software developers.

The (E)MDM includes existence constraints in its mathematical db schemes, along with their duals, the non-existence ones. *MatBase*, an intelligent data and knowledge management system prototype, provides users with a GUI form to declare them, and, if and only if they are well-formed and not violated by the db instance, not only stores them in the corresponding db schemes, but also automatically generates code for their enforcement.

This paper presents and discusses corresponding *MatBase* algorithms, which may also be used by software developers without access to a *MatBase* copy.

As *MatBase* is automatically generating code for constraint enforcement (and not only), it is also a modelware tool [6, 15] and (E)MDM is also a programming language of the 5th generation [15, 16].

## Conflict of Interest

The authors declare that the research was conducted in the absence of any commercial or financial relationships that could be construed as a potential conflict of interest.

## Author Contributions

This paper was written solely by Christian Mancas.

## Funding

This research received no external funding.

## Acknowledgments

This research was not sponsored by anybody and nobody other than its author contributed to it.